\documentclass[12pt,preprint]{aastex631}

\makeatletter
\let\frontmatter@title@above=\relax
\makeatother

\usepackage{CJK}
\usepackage{longtable}
\usepackage{lipsum}
\usepackage{multirow}
\usepackage{natbib}
\usepackage{savesym}
\savesymbol{tablenum}

\usepackage{epsfig}
\usepackage{graphicx}
\usepackage{bm}
\usepackage{amsmath}
\usepackage{amsfonts}
\usepackage{amssymb}
\usepackage{xcolor}
\usepackage{mathrsfs}

\usepackage{CJK}
\usepackage{ulem}
\usepackage{siunitx}
\restoresymbol{SIX}{tablenum}
\usepackage{comment}
\usepackage[mathlines]{lineno}

\def\arcsec{$^{\prime\prime}$}
\def\arcmin{$^{\prime}$}
\def\degree{$^{\circ}$}

\newcommand{\vlsr}{$V_{\mathrm{LSR}}$}

\def\deg{$^{\circ}$}

\newcommand{\kms}{km~s$^{-1}$}

\begin{document}

\title{Detection of a High-Velocity Jet From MWC 349A Traced By Hydrogen Recombination Line Maser Emission}

\shorttitle{}

\author[0000-0002-1082-5589]{Sirina Prasad}\affil{Center for Astrophysics $|$ Harvard \& Smithsonian, 60
  Garden Street, Cambridge, MA 02138, USA}
\author[0000-0003-2384-6589]{Qizhou Zhang}\affil{Center for Astrophysics $|$ Harvard \& Smithsonian, 60
  Garden Street, Cambridge, MA 02138, USA}
\author[0000-0002-3882-4414]{James Moran}\affil{Center for Astrophysics $|$ Harvard \& Smithsonian, 60
  Garden Street, Cambridge, MA 02138, USA}
\author[0000-0002-6368-7570]{Yue Cao}\affil{Center for Astrophysics $|$ Harvard \& Smithsonian, 60 Garden Street, Cambridge, MA 02138, USA}
\author[0000-0003-4493-8714]{Izaskun Jimen{\'e}z-Serra} \affil{Centro de Astrobiolog{\'i}a (CSIC-INTA), Ctra. Ajalvir km 4, Torrej{\'o}n de Ardoz, E-28850, Madrid, Spain}
\author[0000-0003-4561-3508]{Jesus Mart{\'i}n-Pintado} \affil{Centro de Astrobiolog{\'i}a (CSIC-INTA), Ctra. Ajalvir km 4, Torrej{\'o}n de Ardoz, E-28850, Madrid, Spain}
\author[0000-0001-5191-2075]{Antonio Martinez Henares} \affil{Centro de Astrobiolog{\'i}a (CSIC-INTA), Ctra. Ajalvir km 4, Torrej{\'o}n de Ardoz, E-28850, Madrid, Spain}
\author[0000-0001-8125-5993]{Alejandro B{\'a}ez-Rubio} \affil{Mahindra United World College, GHVM+7VM, Paud, Maharashtra 412108, India}

\begin{abstract}
MWC 349A is one of the rare stars known to have hydrogen radio recombination line (RRL) masers. The bright maser emission makes it possible to study the dynamics of the system at milli-arcsecond (mas) precision. We present Atacama Large Millimeter/submillimeter Array (ALMA) observations of the 1.4 mm and  0.8 mm continuum emission of MWC 349A, as well as the H30$\alpha$ and H26$\alpha$ RRLs. Using the most extended array configuration of C43-10 with a maximum baseline of 16.2km, we spatially resolved the H30$\alpha$ line and 1.4mm continuum emission for the first time. In addition to the known H30$\alpha$ and H26$\alpha$ maser emission from a Keplerian disk at LSR velocities from $-12$ to $28$ \kms\ and from an ionized wind for velocities between $-12$ to $-40$ \kms\ and $28$ to $60$ \kms, we found evidence of a jet along the polar axis at \vlsr\ from $-85$ to $-40$ \kms\ and $+60$ to $+100$ \kms. These masers are found in a linear structure nearly aligned with the polar axis of the disk. If these masers lie close to the polar axis, their velocities could be as high as 575 \kms\, which cannot be explained solely by a single expanding wind as proposed in \citet{baez-rubio2013}. We suggest that they originate from a high-velocity jet, likely launched by a magnetohydrodynamic wind. The jet appears to rotate in the same direction as the rotation of the disk. A detailed radiative transfer modeling of these emissions will further elucidate the origin of these masers in the wind.

\end{abstract}

\section{Introduction}
\label{sec:intro}

The emission-line star MWC 349A in the constellation Cygnus 
%($\alpha$ = 20h 32m 45.53s, $\delta$ = $40^{\circ}$39'36.''60) 
is one of the brightest stars in radio continuum emission in the sky \citep{braes1972}. At an estimated distance of 1.2 kpc from the Sun \citep{cohen1985}, it is believed to be a hot, massive B[e] star \citep{merrill1933} with H, He I, and Fe II lines \citep{swings1942}. Due to its peculiar properties, MWC 349A has been extensively studied in optical, infrared, and radio wavelengths. 

The radio continuum emission from MWC 349A exhibits a pinched bipolar geometry resembling an hourglass \citep{white1985}. Its flux and size scale with frequency $\nu$ as $\nu^{0.67}$ and  $\nu^{-0.74}$, respectively \citep{olnon1975, tafoya2004}. The circumstellar disk of MWC 349A lies at the waist of the hourglass shape. The emission above and below the waist is thought to arise from thermal Bremsstrahlung radiation in an ionized wind. For an emitting gas with a uniform electron density $n_e$, its spectral energy distribution (SED) consists of two regimes separated by the turnover frequency: An optical thick part in lower frequencies where fluxes vary as $\nu^{2}$, and an optically thin part at higher frequencies where fluxes vary as $\nu^{-0.1}$. Such SEDs are commonly observed in more evolved HII regions associated with high-mass star forming regions \citep[e.g.,][]{keto2008}. As densities of the ionized gas increase, the turnover frequency shifts to higher values. Since the optical depth $\tau_\nu$ of Bremsstrahlung radiation varies with density and frequency as $n_e^2$ and $\nu^{-2}$, respectively, an object with electron densities decreasing outward can have optical depth $\tau_\nu = 1$ layers occurring at different radii for varying frequencies. At higher frequencies,  the $\tau_\nu = 1$ layer is located at the inner part of the source where the density is higher. This property makes the emission partially optically thick at a range of frequencies, and yields a spectral index of the SED between 2 and -0.1. \citet{olnon1975} demonstrated that a density distribution of $n_e \sim r^{-2.1}$ explains the observed SED of $\nu^{0.7}$ in MWC 349A.

Later studies with higher angular resolutions \citep{white1985, tafoya2004} confirmed that this bipolar structure is consistent with an ionized wind with a neutral disk seen as a dark lane in the infrared wavelengths at its waist \citep{danchi2001}. Such an ionized wind can originate from a photo-evaporating, rotating disk \citep{hollenbach1994, danchi2001, hamann1988}, as confirmed by the detection of rotation in the ionized envelope surrounding MWC 349A \citep{rodriguez1994}. The typical velocity of the photo-evaporating wind is comparable to the sound speed of the ionized gas, which is 10 to 20 \kms. The presence of a density gradient in the medium can accelerate the wind to several tens of \kms.  This mechanism may explain the low velocity wind up to 50 \kms\ in MWC 349A \citep{aitken1990}, which is much slower than the typical stellar wind of massive stars \citep{altenhoff1981}.

In addition to strong continuum emission, MWC 349A was the first of the few hydrogen radio recombination line (RRL) masers detected \citep{martin1989a, strelnitski1996b}. \citet{martin1989a} reported strong and double-horn spectra in RRLs H31$\alpha$, H30$\alpha$ and H29$\alpha$. The line-to-continuum ratios of these transitions are much higher than the value expected under a local thermodynamic equilibrium (LTE). Therefore, these emissions are masers (microwave amplification by stimulated emission of radiation)  \citep{martin1989a}. Subsequently, maser emission was found in other $\alpha$ transitions down to the principle quantum number $n = 7$ \citep{martin1994, thum1998}.

Maser emission requires a velocity coherence and a level population inversion in the emitting medium \citep{elitzur1989, strelnitski1996a}. As an electron combines with an $H^+$ and cascades down to lower energy levels, the difference in Einstein A coefficients for the $\alpha$ transitions, i.e., $A_{n, n-1}$, favors an over population of the upper energy levels. Therefore, emission of hydrogen RRLs is often weakly inverted \citep[]{jimenez2013}. However, high amplifications as exhibited in the RRLs in MWC 349A remain rare despite the wide-spread population inversion observed in HII regions. This is in contrast to the copious H$_2$O masers found in envelopes of both young and evolved stars. 

Interferometric observations of MWC 349A have been very useful in probing the spatial distribution of the maser emission. The first such observations, from the Owens Valley Millimeter Interferometer, revealed an east-west double-peaked structure in the H30$\alpha$ maser emission \citep{planesas1992}. Further observations from the PdBI and the SMA, whose high signal-to-noise ratios in the maser emission enable the determination of the precise positions of the velocity components by use of using the centroid fitting techniques, provided strong evidence that the masers arise from a nearly edge-on Keplerian disk \citep{Condon1997,weintroub2008,martin2011,zhang2017}. The most accurate observations thus far show H30$\alpha$ masers in a linear distribution in the southeast and northwest direction with an extent of 45 mas and a position angle (PA) of $101^\circ$, for emissions between V$_{LSR}$ velocities of -12 to 25 \kms. The H26$\alpha$ emission observed with the SMA revealed a linear distribution in agreement with those of the H30$\alpha$ masers, but with a spatial extent of only 40 mas inside of the H30$\alpha$ masers \citep{zhang2017}. Maser emissions outside of the velocity range of -12 to 25 \kms\ arise from a rotating and expanding wind \citep{rubio2014}, although the mechanism and dynamics behind the ejection of the ionized wind remain elusive. Various models, such as photo-evaporating wind \citep{hollenbach1994}, a spiral structure in the envelope of MWC 349A \citep{weintroub2008}, and a magneto-hydrodynamic wind \citep{zhang2017, martin2011, baez-rubio2013} have been proposed.

While the SMA and PdBI observations have provided valuable insights into the spatial distribution and kinematics of the masers in the disk and the wind, these studies are limited by their angular resolutions and sensitivities which were not sufficient to either spatially resolve or obtain accurate positions for the emission, especially for the high velocity features in the line wings beyond $\pm$40~\kms\ from the systemic velocity. Furthermore, the relative position accuracy between masers in different velocities is limited by the phase noise in the passband calibration. For the SMA observations, \citet{zhang2017} found a $1 \sigma$ phase noise of $2^\circ$ in the passband calibration, which translates to position errors of 2.5 and 1.8 mas for the H30$\alpha$ and H26$\alpha$ masers, respectively. With a much larger collecting area and long array baselines up to 16 km,
the Atacama Large Millimeter/submillimeter Array (ALMA) is capable of attaining much higher angular resolutions and sensitivities than the SMA and PdBI, and therefore, has the potential to produce more accurate or even spatially resolved maser distributions over a wider range of radial velocities. Here, we present ALMA observations of H26$\alpha$ and H30$\alpha$ RRL masers in MWC 349A at angular resolutions of 0.1\arcsec\ and 0.03\arcsec, respectively. The new observations yield the first spatially resolved continuum emission of MWC 349A at 1.4mm, and allowed for the detection of higher velocity maser emissions. We apply the position centroid method to obtain detailed maser distributions and rotation curves. The sensitive and high-angular resolution observations led to the detection of a new high velocity wind component close to the rotation axis of the disk. 

The paper is organized as follows. In Section \ref{sec:obs}, we present detailed information on the observations, data calibration, and imaging. Section \ref{sec:results} presents the main results and discussions. We conclude with a summary of the main findings in Section \ref{sec:summ}.

\section{Observations}
\label{sec:obs}

\subsection{ALMA}
We carried out observations of MWC 349A using the Atacama Large Millimeter/submillimeter Array (Project ID: 2017.1.00404.S; PI: Q. Zhang) in band 6 and band 7, with frequency ranges of 211 to 275 GHz and 275 to 370 GHz, respectively. The band 6 observations were executed on 2017 October 5 in the array configuration of C43-10, while the band 7 observations were obtained on 2019 August 24 using the array configuration of C43-7. The ALMA correlator was configured to observe the H30$\alpha$ line (rest frequency of 231.9009 GHz) in band 6 and the H26$\alpha$ line (rest frequency of 353.6228 GHz) in band 7, with four spectral windows in each band. Two spectral windows were tuned to the hydrogen recombination line frequency using a bandwidth of 1.875 GHz and 0.4688 GHz, respectively, and 1920 spectral channels for each window, of width 0.9766 MHz and 0.2441 MHz, respectively. The remaining two windows were used for continuum measurements in the Time Division Mode (TDM) with a bandwidth of 1.875 GHz and 128 channels. The phase center of the MWC~349A observations was RA (J2000) = ${\rm 20^h32^m45^s.5280}$, Dec(J2000)= 40\degree39\arcmin36\arcsec.623. For the band 6 observations, quasar J2148+0657 was used as the bandpass calibrator, and J2007+4029 as the time dependent gain calibrator. For the band 7 observations, quasar J2253+1608 was used as the bandpass calibrator, and J2015+3710 was adopted for time dependent gain calibration. Calibrations of visibilities were performed by the ALMA support staff using the Common Astronomy Software Applications package (CASA) \citep{mcMullin2007}. The calibrated measurement set of MWC~349A was further self calibrated by use of continuum data.

We imaged the band 6 and band 7 continuum emission and the H30$\alpha$ and H26$\alpha$ lines of MWC 349A using CLEAN-based algorithms in CASA. For both band 6 and band 7, the continuum emission was constructed from the channel-averaged combination of the two 1.875GHz spectral windows with frequency ranges lower than those of the H30$\alpha$ and H26$\alpha$ recombination lines, respectively. In order to improve the dynamical range of the images, we performed self calibrations of the continuum data in each band and apply the solutions to the corresponding line data. Self calibrations used a 30s time interval to solve for short-term gain variations. We performed three iterations in gain phases, and the fourth iteration in both phase and amplitude. We then imaged the self-calibrated continuum visibilities using the Briggs weighting with a robustness parameter of -2. The synthesized beam for the band 6 continuum image is 0.075\arcsec $\times$ 0.025\arcsec, with a position angle of 4.4\degree, and for band 7 it is 0.107\arcsec $\times$ 0.051\arcsec, with a position angle of -5.3\degree.

The solutions from the self-calibration of the continuum data  were applied to the spectral windows containing the H30$\alpha$ and H26$\alpha$ lines, which were then imaged by use of natural weighting to optimize signal-to-noise ratios of the line emission. The H30$\alpha$ data were channel-averaged to match the spectral resolution of the H26$\alpha$ line of 1.2 \kms. Local Standard of Rest (LSR) radio velocity was calculated using the rest frequencies for H30$\alpha$ and H26$\alpha$, respectively, and the standard International Astronomical Union model. The self calibration improves the dynamical ranges in both continuum and the recombination line images by more than one order of magnitude. The final images reached a dynamic range of $\sim 10^3$. Since self calibration removes information about absolute position, images in band 6 and 7 lost their absolute astrometry. In order to compare the spatial distribution of the H30$\alpha$ and H26$\alpha$ emission, we register the line emission relative to the continuum image in their respective band. Since the continuum emission in band 6 and 7 traces nearly the same material, this allowed comparisons of maser emission from the two lines at a high positional accuracy.

\subsection{SMA}
Observations of MWC 349A with the Submillimeter Array (SMA)\footnote{The SMA is a joint project between the Smithsonian Astrophysical Observatory and the Academia Sinica Institute of Astronomy and Astrophysics, funded by the Smithsonian Institution and the Academia Sinica.} \citep{ho2004} were carried out on 2014 August 14 with eight antennas in its very extended configuration with a maximum baseline of 0.5km. Both 230GHz and 400GHz receivers were used in the Array and were tuned to the H${30 \alpha}$ and H${26 \alpha}$ lines, respectively. With an IF frequencies of 4 to 6 GHz, the total spectral bandwidth was 4 GHz for each frequency band. The correlator was configured to 256 channels per 104-MHz window for three consecutive spectral windows around the recombination lines. The remaining 21 windows in the 2-GHz IF band were set to 32 channels per 104-MHz spectral window.

Quasar J2015+371 was observed periodically to monitor gain variations during the course of the observations. In addition, Titan and Neptune were observed as flux calibrators  and 3C454.3 and 3C84 were observed as passband calibrators. Calibrations were performed using the IDL subset MIR software package. We found a flux density of 1.66 Jy at 230 GHz and 2.35 Jy at 345 GHz for the gain calibrator J2015+371. With a maximum baseline of 509 m, MWC 349A was unresolved in both recombination line and continuum emission. The measured continuum flux for MWC 349A was 1.7 Jy at 221.9 GHz, and 2.4 Jy at 345.6 GHz, respectively. 

We also make use of additional SMA archival data of MWC 349A obtained on 2012 October 12 in very extended configuration. These observations used six antennas in the Array. The correlator setup was identical to those in the 2014 August 14 observations described above. Additional details on observations and data calibration can be found in \citet{zhang2017}. The measured continuum flux density for MWC 349A was 1.7 Jy at 221.9 and 2.4 Jy at 343.6 GHz, respectively. The typical uncertainty in flux calibrations with the SMA is 10\% for 230 GHz observations and 15\% for 345 GHz observations.

\section{Results and Discussion}
\label{sec:results}

\subsection{Interferometric Continuum Emission}

% band 7 Gaussian fit: sigma_x = 0.048'' and sigma_y = 0.062''
% band 6 Gaussian fit - sigma = 25 mas. 

\begin{figure}
    \centering
    \includegraphics[width=\linewidth]{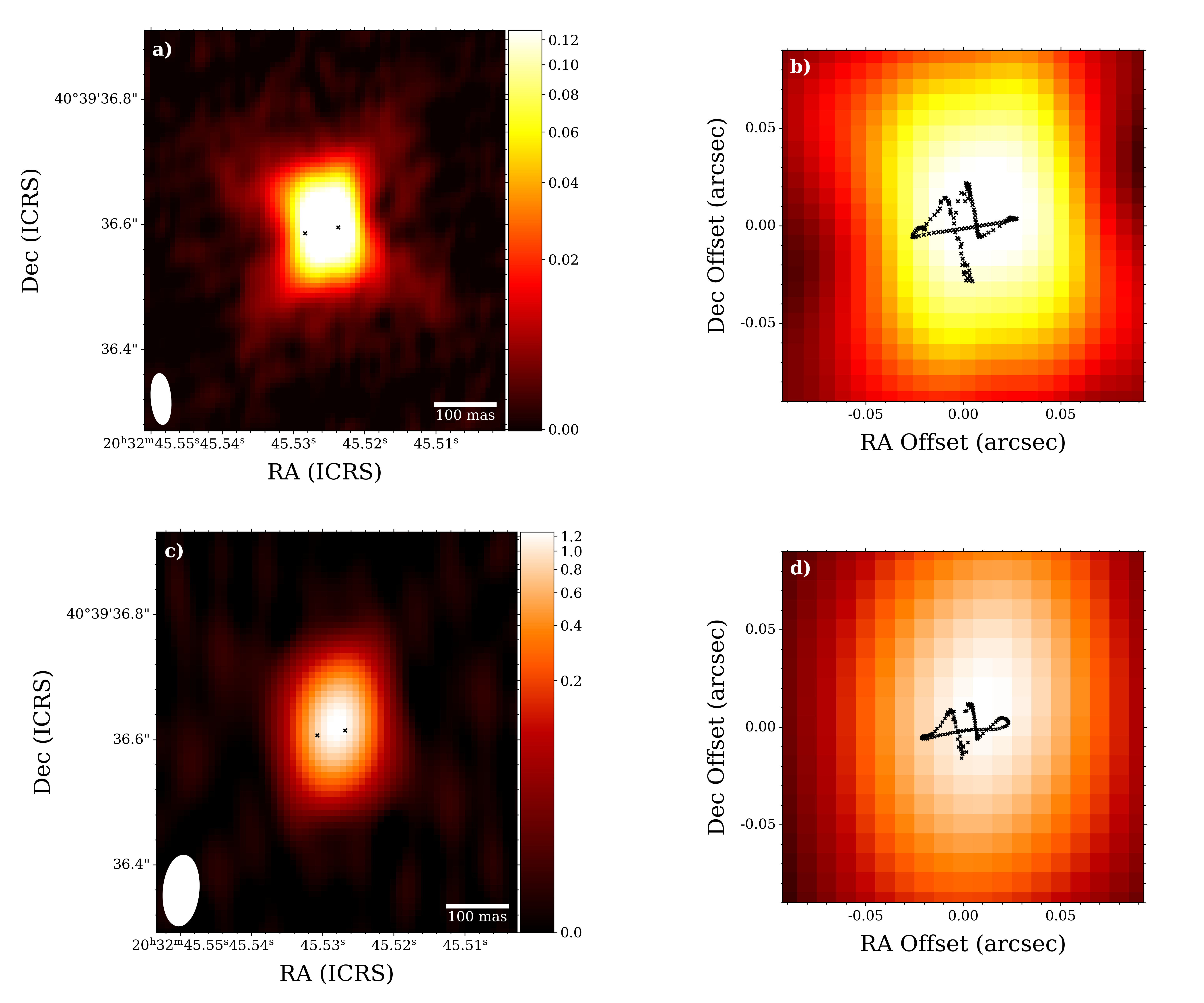}
    \caption{Continuum emission of MWC 349A at 1.4mm in Band 6 (panel a) and 0.87mm in Band 7 (panel c) obtained with ALMA. Color indicates intensities in Jy Beam$^{-1}$, and the ellipse at the lower-left corner marks the synthesized beam. The two overlaid points in panels (a) and (c) correspond to the two ends of the disk as seen in the maser distribution for that band. Panels (b) and (d) are zoomed-in areas of panels (a) and (c), respectively.}
    \label{fig:continuum}
\end{figure}

Figure \ref{fig:continuum} presents images of the continuum emission from MWC 349A in the ALMA band 6 and band 7. The H30$\alpha$ and H26$\alpha$ masers (see Section 3.2) are also overlaid on the continuum images in Figures \ref{fig:continuum}(b) and \ref{fig:continuum}(d), respectively. The frequency range was $215.00 - 219.00$ GHz for the band 6 continuum, and $339.61 - 343.51$ GHz for the band 7 continuum emission. In both bands, the continuum is taken at a frequency slightly lower than the hydrogen RRLs. The band 7 continuum image was modeled well by an elliptical Gaussian centered at RA $ = {\rm 20^h32^m45^s.53}$, Dec = +40\degree39\arcmin36\arcsec.54, with position angle -6.32\degree\, and (at FWHM) a major axis of size 0.125\arcsec\ and minor axis of size 0.080\arcsec. Note that the synthesized beam (0.107\arcsec $\times$ 0.051\arcsec) is nearly in the north-south direction, thus this image essentially shows the MWC 349A system as a point source. The deconvolved size, obtained by quadratically subtracting the beam size from the FWHM of the Gaussian fit, is 0.10\arcsec $\times$ 0.10\arcsec. In band 6, with a beam size of 0.075\arcsec $\times$ 0.025\arcsec, the continuum emission was spatially resolved, revealing a bipolar morphology with a waist. This hourglass shape is particularly visible in the low level emission on the edges. An elliptical Gaussian fit of this image yields an estimate for the center of the band 6 emission: RA = ${\rm 20^h32^m45^s.53}$, Dec = +40\degree39\arcmin36\arcsec.59. The approximate extent of the hourglass structure, as measured by the FWHM of this Gaussian, is 59 mas in the east-west direction. The difference in the continuum positions of the two bands likely arise from the self calibration discussed in Section \ref{sec:obs}.

The continuum images from ALMA are in good agreement with previous observations at longer wavelengths. The resolved morphology of the 1.4mm continuum emission is in agreement with the data obtained with the VLA at 2, 1.3 and 0.7 cm \citep{white1985, tafoya2004}. At all these wavelengths, the continuum emission exhibits an hourglass morphology with a slightly-angled waist \citep{martin1993}. Others \citep{rodriguez2007,baez-rubio2013} have estimated this inclination angle to be around 98\degree\ in PA; our 1.4mm continuum image above appears to be consistent with this finding. Unlike previous cm continuum observations, however, we did not detect a “dark lane” separating two lobes in an approximate north-south direction \citep{tafoya2004, white1985}. Both images from ALMA consist of a single peak. This is most likely a result of a relatively large beam size elongated in the north-south direction, which was insufficient to resolve the double peaks aligned in the same direction. 

\begin{table}
    \centering
    \begin{tabular}{c|c|c}
        \text{Frequency (GHz)} & \text{Total Flux Density (mJy)} & \text{Angular Size (FWHM) (\arcsec)} \\
        \hline \hline
        217.0 & 1620.8 & 0.0718 \\
        \hline
        341.6 & 2754.6 & 0.0421
    \end{tabular}
    \caption{Total flux and angular sizes for ALMA band 6 ($217.0$ GHz) and band 7 ($341.55$ GHz) continuum images.}
    \label{tab:fluxsize}
\end{table}

\begin{figure}
    \centering
    \includegraphics[width=0.9\textwidth]{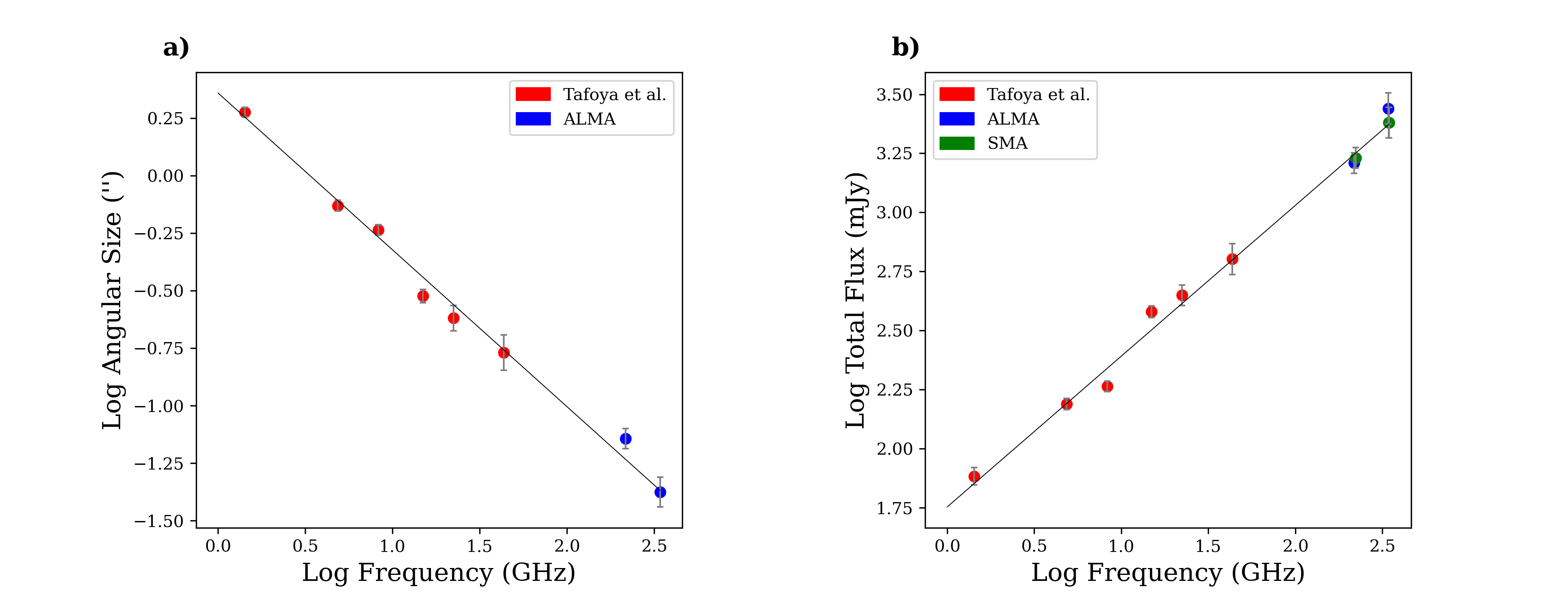}
    \caption{Total flux density and angular size as a function of frequency for the continuum emission of MWC 349A. The solid lines are the linear least-squares fits to the data from this study and data from previous studies; the dashed lines are the linear least-squares fits found by \cite{tafoya2004}.}
    \label{fig:scalingrelations}
\end{figure}

The flux densities of the continuum emission also reinforce the trend in frequency found in the previous observations of MWC 349A. The most recent analysis of these trends, done by \cite{tafoya2004}, is in excellent agreement with relations found by others (e.g., \cite{dreher1983}, \cite{altenhoff1981}). As formulated by \cite{escalante1989}, the angular size and total flux density can be computed directly from $u$-$v$ data using the relation 
\begin{equation}
    V(b) = S_\nu(1-Ab)
\end{equation}
for a spherically symmetric ionized wind source extending to infinity, where $S_\nu$ is the total flux density, $A$ is a constant, and $b$ is the projected baseline distance. As this linear relation holds only for short baselines, a linear regression was performed only on the visibility data with baseline less than $500,000$ wavenumbers. This was done for data from both ALMA bands to obtain estimates for the total flux densities. Estimates for the angular size were then calculated using the following relation \citep{tafoya2004}
\begin{equation}
    \Theta \text{/arcsec} = 1.19 \times 10^5 A.
\end{equation}
The total flux and angular size derived from our continuum data are listed in Table \ref{tab:fluxsize}. Figure \ref{fig:scalingrelations} shows the angular size and the flux versus frequency, including the data points reported in \cite{tafoya2004}. The flux density was also measured with the SMA between 2012 and 2014, resulting in several more points on the total flux plot. A least-squares power-law regression with the addition of these new points yields scaling relations:
\begin{align}
    \Theta_\nu\text{/arcsec} &= (2.3\pm 0.2)(\nu\text{/GHz})^{-0.68 \pm 0.03}, \\
    S_v\text{/mJy} &= (56 \pm 4)(\nu\text{/GHz})^{0.64 \pm 0.02}.
\end{align}
These relations are remarkably similar to those found by \citet{tafoya2004}. In particular, the new data points provide evidence that the previously observed scaling relations continue to higher frequencies, indicating that the continuum emission is still partially optically thick even at 340 GHz.

\subsection{Hydrogen Recombination Line Emission}

\begin{figure}
    \centering
    \includegraphics[width=0.49\linewidth]{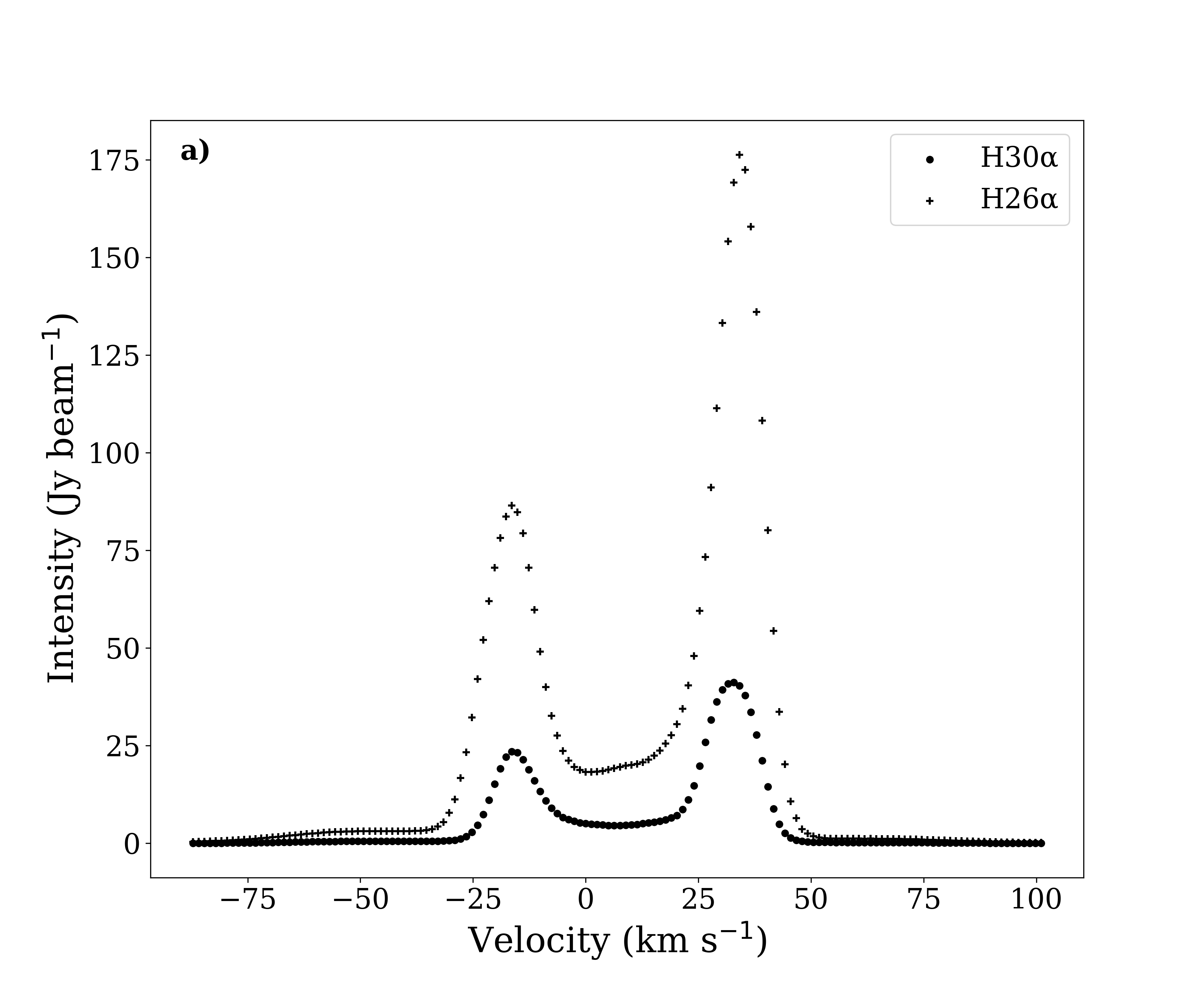}
    \includegraphics[width=0.49\linewidth]{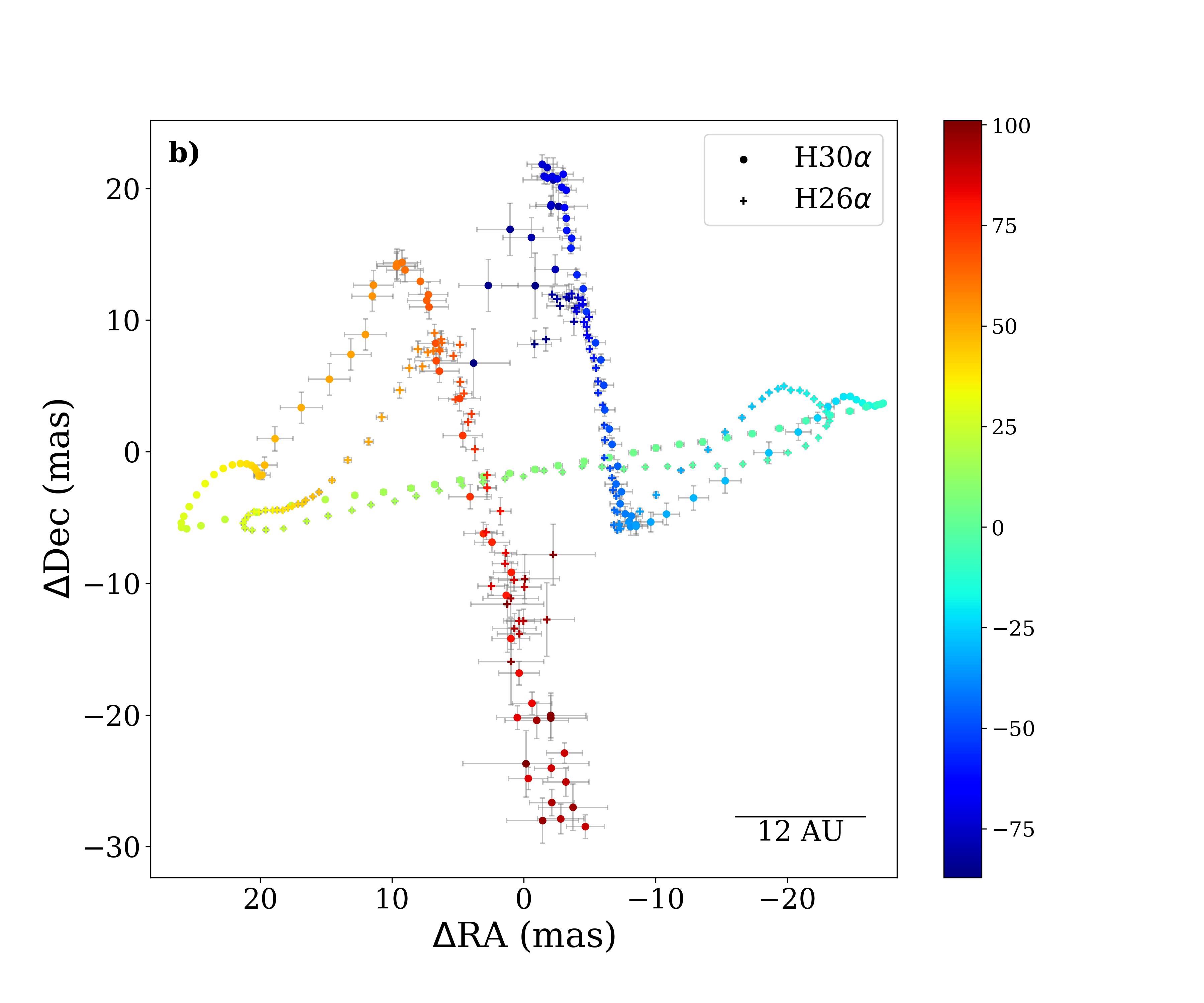}
    \caption{\textbf{(a)} On the left, the spectral profile plot for the H26$\alpha$ and H30$\alpha$ lines. The peak intensity for each spectral channel in the band 6 and 7 data cubes was extracted as the amplitude from the centroid fit. Velocities are measured in the LSR reference frame in km~s$^{-1}$, and for both lines the velocitry resolution is $1.26$ \kms. \textbf{(b)} On the right, the spatial distributions of H30$\alpha$ and H26$\alpha$ maser emission obtained by fitting elliptical Gaussians to the data cubes for the two recombination lines. Error bars include the standard deviation from the centroid fit and error contribution from the bandpass calibration (see text for details). The H26$\alpha$ and the H30$\alpha$ distributions are offset from the position of the band 7 and band 6 continuum, respectively.}
    \label{fig:spectralprofile}
    \label{fig:maserdist}
\end{figure}

The positions and intensities of the H26$\alpha$ and H30$\alpha$ maser emission as a function of radial velocity were obtained from the spectral line data cubes by use of a centroid fitting method. A 2-dimensional Gaussian function was fitted to the image for each spectral channel using the Levenberg-Marquardt least-square algorithm, yielding the central position, the major axis, minor axis, and the position angle of the ellipse, and the peak amplitude. The amplitude of the fit is taken to be the intensity (in Jy/beam) of the maser. The resulting spectral profiles are shown in Figure \ref{fig:spectralprofile}\textbf{(a)}. For both lines, the red-shifted peak is stronger than the blue-shifted peak. For all velocities, the H26$\alpha$ line is much stronger than the H30$\alpha$ line. This is partially due to the fact that the H30$\alpha$ line is observed with a beam area a factor of 3 smaller than that of the H26$\alpha$ line, indicating that the maser emission is spatially extended.

To obtain the spatial distribution of masers, the offset of the central position of the fit for each channel was plotted with respect to the center of the continuum emission as determined by the elliptical Gaussian fit. To account for a position offset between the two bands caused by self-calibration, we computed relative positions for masers with respect to the centroid position of the band 6 and band 7 continuum emission, respectively. The resulting distribution of masers is shown in Figure \ref{fig:maserdist}(b). Each point is color coded by the LSR velocity of the channel image. The positional errors in the maser distribution are two-fold: The error of the centroid fit, and the error associated with the bandpass calibration.

% This suggests that the maser emission from this system is well-modeled by point sources.

A reliable determination of the spatial distributions of masers also requires an accurate bandpass calibration, since phase errors in the spectral bandpass solutions are added to the line data of MWC~349A when bandpass calibrations are applied. In order to calculate the phase errors, we estimated the signal to noise ratios from the bandpass calibrators employed for the observations. For J2148+0657 used in the band 6 observations, we found a SNR of 470 in images with a channel width of 1.26 \kms. This corresponds to a position error of 0.042 mas in band 6. J2253+161 was used as the bandpass calibrator in the band 7 observations. We found a SNR of 1400 in images with a channel width of 1.26 \kms, and a corresponding error of 0.038 mas. These errors were quadratically added to the fit errors in order to obtain the error bars in Figure \ref{fig:maserdist}(b).

As shown in Figure \ref{fig:maserdist}(b), masers with velocities between -12~\kms\ and 28~\kms\ are approximately distributed in a linear structure. These masers arise from a disk viewed nearly edge on, as reported by \citet{weintroub2008, martin2011, zhang2017}. The masers have a maximum spatial extent of approximately $45$ mas in the H$26\alpha$ transition and $54$ mas in the H$30\alpha$ transition. Previous data have supported a model in which the density of the disk has only a radial dependence and the two maser transitions are localized on two annuli of different radii \citep{zhang2017}. With this model, the differing spatial extents of H$26\alpha$ and H$30\alpha$ in these data indicate an inverse relationship between density and radius, as is typically expected for circumstellar disks.

A linear least-squares fit to these maser points in the disk gives a position angle of the disk axis of $98.6\pm0.007$\degree\ in H$26\alpha$ and $99.7\pm0.002$\degree\ in H$30\alpha$. On the blue shifted side, the masers rise above the plane of the disk and then dip below it, continuing the downward trend until -40~\kms. On the red shifted side, the masers rise above the disk until 60~\kms. At extreme velocities ($\geq$ 60~\kms\ and $\leq$ -40~\kms), our ALMA images reveal for the first time that the masers lie along lines perpendicular to the plane of the disk, very close to the polar axis of the system. The red shifted masers appear to shift downward toward the south and away from the system. Similarly, the blue shifted masers appear to shift upward toward the north and away from the system. For both blue-shifted and redshifted masers, the H$30\alpha$ emission extends slightly further out from the disk than the H$26\alpha$ emission. At the most extreme velocities of $-80$ and $+90$ \kms, both blue and red shifted masers appear to shift toward the center of the disk along the polar axis.

The limited signal-to-noise ratios in the high velocity line wings of previous observations and studies of distributions of MWC 349A hydrogen masers have prevented a precise determination of the position and kinematics of the maser emission in the wind and outflow. The measurement of the spatial distribution of these maser emissions was made possible by the superb sensitivity and the high angular resolution of the ALMA images. Figure \ref{fig:maserdist}(b) exhibits several noteworthy new features: (1) The blue-shifted masers appear to rise above the disk at the very edge, whereas on the opposite side, the red-shifted masers do not dip below the disk.
(2) The highest velocity blue-shifted masers rise up along the axis of rotation but then begin to move downward, back towards the disk. (3) The H26$\alpha$ redshifted masers display a similar motion: they move downwards along the axis of rotation but then begin moving upwards towards the disk. The H30$\alpha$ red-shifted masers, on the other hand, move down and away from the star-disk system along its axis of rotation, and then back toward the disk plane at the most extreme velocities.

\begin{figure}
    \centering
    \includegraphics[width=0.9\linewidth]{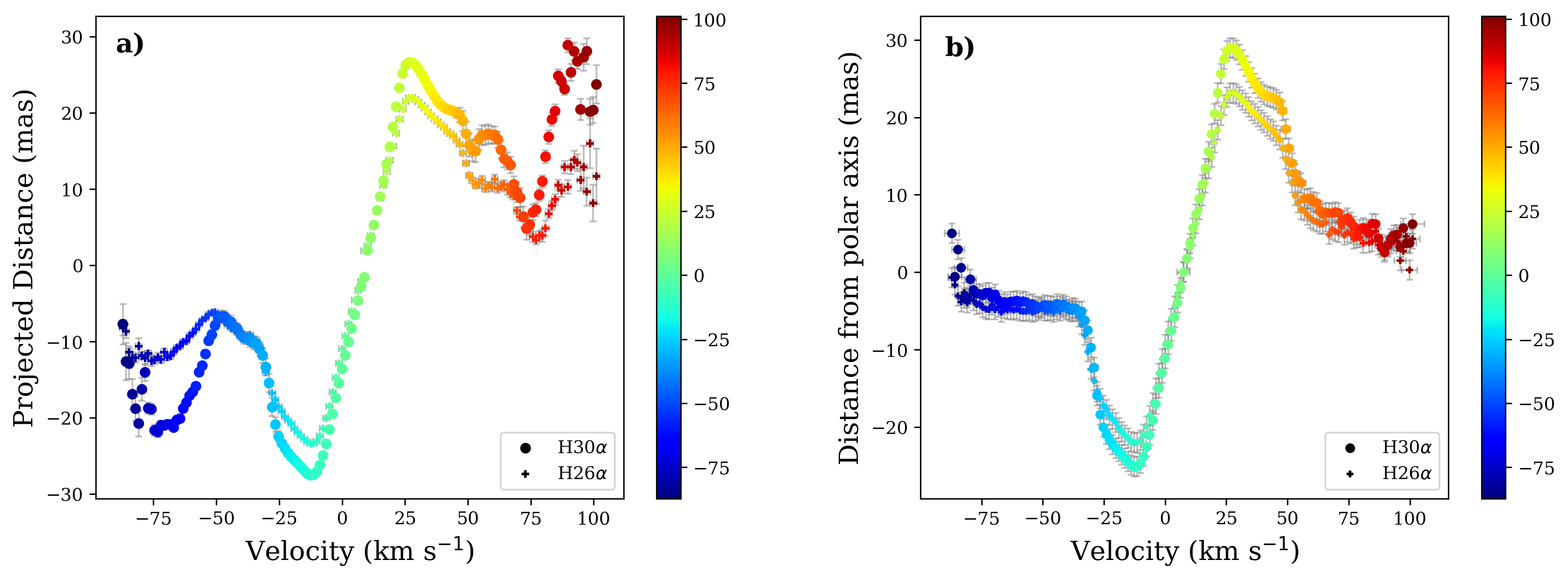}
    \caption{Rotation curves for H30$\alpha$ and H26$\alpha$ masers in MWC 349A. On the left, the projected distance is calculated as $\sqrt{(\Delta \text{RA} \cos{(\text{Dec}}))^2 + \Delta \text{Dec}^2}$. On the right, distance to the polar axis is calculated as the perpendicular distance from the maser point to the line that is both orthogonal to the line of best fit for the disk masers and that intersects the best fit line at the center. In both plots the color indicates the velocity of the masers in km s$^{-1}$.}
    \label{fig:velcurve}
\end{figure}

To better understand the kinematics of the masers, we used the data from the centroid fitting to produce two types of position-velocity curves for both lines, shown in Figure \ref{fig:velcurve}. On the left panel, the projected distance is measured from each maser point to the center of the disk, the position of the continuum emission in each band. The error in the projected distance was carried over from the error shown in Figure \ref{fig:maserdist}(b); the error in velocity is assumed to be the width of the spectral channel. For both transitions, one finds a linear trend between -12 \kms\ and 28 \kms. The y-intercept of a linear fit to the disk portion of the curve gave an estimate for the systemic velocity of this star-disk system. Using the H30$\alpha$ rotation curve, this linear fit yielded a y-intercept of 8.7 ± 0.2 \kms, and using the H26$\alpha$ rotation curve, we obtain 8.4 ± 0.1 km/s. 

The slopes obtained from these linear fits to the disk maser position-velocity plots were 771 ± 7 km s$^{-1}$ per arcsecond and 661 ± 9 km s$^{-1}$ per arcsecond for H$26\alpha$ and H$30\alpha$ masers, respectively. Previously, \cite{zhang2017} assumed a model in which the two maser transitions are localized on two thin annuli of different, fixed radii on a perfectly edge-on disk. In this case, the velocity-projected distance slopes can be described by the following relation (assuming Keplerian motion):
\begin{equation}
    V_{LOS} = \sqrt{\frac{GM}{R^3}}l
\end{equation}
where $V_{LOS}$ is the line-of-sight (LOS) velocity, $l$ is the projected distance, and $R$ is the radius of the annulus. For each transition, we took the radius of the annulus to be the spatial extent of the masers; the slopes from the linear fit can then be used to obtain an estimate of the enclosed mass. The H$26\alpha$ masers yield an estimate of 9.6 ± 1.6 M$_\odot$, while for the H$30\alpha$ masers it is 11.5 ± 1.7 M$_\odot$. This is in excellent agreement with previous estimates, which put the mass of this system at $10-15$ M$_\odot$ \citep{zhang2017}. These estimates of the stellar mass are smaller than the model value of 38 M$_\odot$ in \citet{baez-rubio2013}. This is mainly due to the fact that the size of the maser disk is underestimated with the centroid fit since the maser emission is spatially extended as shown in the model by \citet{baez-rubio2013}.

In Figure \ref{fig:velcurve}(b), the distance from the polar axis was calculated as the shortest distance from the maser to the line representing the polar axis of the disk, where the polar axis line is orthogonal to the best fit line to the masers in the disk-like linear structure and intersects with the center of the disk. We found trends similar to those in the plot on the left, but here the red- and blue-shifted sides were more clearly differentiated. The blue-shifted masers show a nearly flat-line trend at extreme velocities ($v<-30$ \kms), meaning the masers maintain their distance from the polar axis as their velocities become more blue shifted. The red-shifted masers, on the other hand, exhibit a small bump in the mid-velocity range before slowly moving towards the polar axis as velocity increases. Of note is the fact that, at extreme blue and red shifted velocities, the masers approach the polar axis. 

\begin{figure}
\begin{interactive}{js}{interactive.zip}
    \centering
    {{\includegraphics[width=0.45\linewidth]{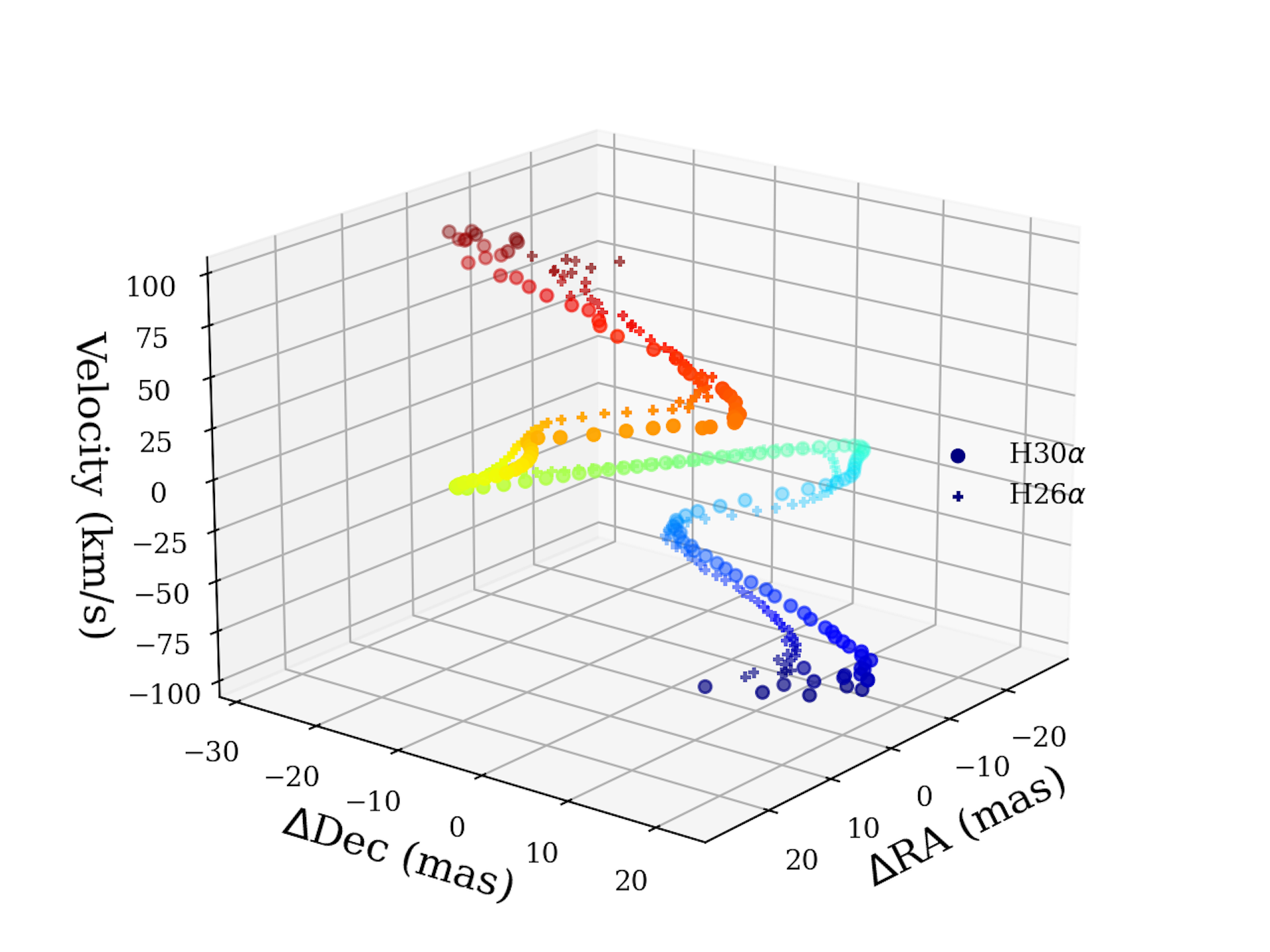} }}
    \caption{A 3D visualization of the H30$\alpha$ and H26$\alpha$ maser distributions, as a function of line-of-sight velocity and 2-dimensional spatial coordinates. This figure is available online as an interactive figure.}
    \label{fig:ppv}
\end{interactive}
\end{figure}

An interactive 3D visualization of the position and velocity of these masers is shown in Figure \ref{fig:ppv}. The high velocity maser features identified in the ALMA observations have line-of-sight velocities from $\pm$50 to $\pm$90 \kms\ from the systemic velocity. Since the disk is nearly edge-on with an inclination angle of 9\deg\ from the LOS, the space velocity of these maser features can be as high as $V_{LOS}/$sin(9\deg), or up to 575 \kms, if these masers are distributed along the polar axis. On the other hand, the expected velocity of a photo-evaporating wind, calculated as the escape velocity for the radius at which the wind is launched, is typically comparable to the sound speed \citep{hollenbach1994}.
Therefore, these masers, with velocities more than one order of magnitude higher than the expected velocity, cannot be explained by a photo-evaporating wind. The fact that they are distributed close to the rotation axis of the disk suggests the presence of a high-velocity jet component in the wind.

The origin of the high velocity jet in MWC349A is not clear. A likely candidate is the magnetohydrodynamic wind \citep{konigl2000, shang2007} widely observed in protostars \citep[][]{palau2006, lee2007a, lee2015}, but it has also been reported in evolved stars \citep{lee2013}. The wind is launched off the disk at locations where the magnetic field is pinched, with a small opening angle \citep{shu1999}. It has been shown that a jet and a wide angle wind are both present in protostellar outflows \citep{shang2006, torrelles2011}, a scenario that may also apply to the MWC 349A system, although MWC 349A is likely an evolved supergiant star. The double-horned spectral profile seen in the masers, spanning both negative and positive velocity ranges, indicate rotation in the wind and the jet. A detailed radiative transfer modeling of the maser emission is underway (Martinez-Henares et al. in prep), and will help to understand the origin of these masers with respect to the wind and the jet. An order of magnitude improvements in sensitivity over the SMA and PdBI observations allowed us to discover the existence of the jet; more sensitive observations of even higher velocity features will help elucidate the nature of this jet.

\section{Summary}
\label{sec:summ}

We have presented the analysis of the H30$\alpha$ and H26$\alpha$ radio recombination lines as well as 1.4mm and 0.87mm continuum emission of MWC~349A based on ALMA observations in array configurations C43-10 and C43-7, respectively. The main findings are as follows:

The 1.4mm continuum emission is spatially resolved in an hourglass morphology for the first time with the most extended array configuration. Combining with the spatially resolved images of MWC~349A obtained with the VLA, we find that the angular size (corresponding to the FWHM for the $\tau=1$ surface) of the continuum emission follows the relation: $\Theta_\nu\text{/arcsec} = (2.3\pm 0.2)(\nu\text{/GHz})^{-0.68 \pm 0.03}$.

The high sensitivity and high angular resolution data of the recombination line images have allowed us to determine accurate positions of masers of LSR velocities from $-$85 to 100 \kms. While masers from $-$12 to 28 \kms\ are distributed in a linear structure arising from a Keplerian disk, those outside of this velocity range are offset from the disk arising from the ionized wind. We identified a new group of masers with extreme redshifted and blue-shifted velocities that are distributed close to the polar axis of the disk. These masers could have space velocities exceeding 575 \kms\ after correcting for the projection effect. Such high velocities cannot be explained by a photo evaporated wind, and require magnetically driven wind seen in both young and evolved stars.
\\ \hfill \\
\hfill \\
This paper uses the following ALMA data: ADS/JAO. ALMA No. 2017.1.00404.S. ALMA is a partnership of ESO (representing its member states), NSF (USA) and NINS (Japan), together with NRC (Canada), $MOST$ and ASIAA (Taiwan), and KASI (Republic of Korea), in cooperation with the Republic of Chile. The Joint ALMA Observatory is operated by ESO, AUI/NRAO, and NAOJ.
The National Radio Astronomy Observatory is a facility of the National Science Foundation operated under cooperative agreement by Associated Universities, Inc.

S.P. and Q.Z. acknowledge NRAO SOS support. I.J-.S, J.M.-P. and A.M.-H. acknowledge funding from grant No. 
PID2019-105552RB-C41 awarded by the Spanish Ministry of Science and 
Innovation/State Agency of Research MCIN/AEI/10.13039/501100011033.
A.M.-H. has received support from grant MDM-2017-0737 Unidad de Excelencia ``María de Maeztu" Centro de Astrobiología (CAB, CSIC-INTA) funded by MCIN/AEI/10.13039/501100011033.

\bibliography{bibliography.bib}

\end{document}